\documentclass[a4paper,11pt]{article}
\usepackage{pos}
\usepackage{comment}

\usepackage{amsmath}
\usepackage{bm}

\newcommand{\Jpsi}{J/\psi}
\newcommand{\qT}{\bm{q}_T}

\newcommand{\Kperp}{\bm{K}_\perp}
\newcommand{\MQ}{M_{\cal Q}}
\newcommand{\Mperp}{M_\perp}
\newcommand{\Cc}[1]{{\cal C}\!\left[#1\right]}

\newcommand{\fperpT}{f_{1T}^{\perp}}
\newcommand{\avg}[1]{\langle #1 \rangle}
\newcommand{\qqbar}{q\bar q}
\newcommand{\sqrts}{\sqrt{s}}

\title{Double quarkonium production in hadronic collisions at fixed-target experiments}
\ShortTitle{Double quarkonium production at fixed-target experiments}

\author*[a,b]{Carlo Flore}
\author[a,b]{Cristian Pisano}

\affiliation[a]{Dipartimento di Fisica, Universit\`a di Cagliari,\\
Cittadella Universitaria, I-09042 Monserrato (CA), Italy}
\affiliation[b]{INFN, Sezione di Cagliari,\\
Cittadella Universitaria, I-09042 Monserrato (CA), Italy}

\emailAdd{carlo.flore@unica.it}
\emailAdd{cristian.pisano@unica.it}

\abstract{We present new results for double quarkonium production in (un)polarized hadronic collisions at fixed-target experiments. Our approach incorporates the transverse momentum dependent factorization in combination with the Color-Singlet Model. We present new analytical expressions for the angular structure of the cross section for the $q\bar q$-induced channel, and provide predictions for the unpolarized cross section and transverse single-spin asymmetries for present and future fixed-target experiments at CERN and the LHC.}

\FullConference{The 33rd International Workshop on Deep Inelastic Scattering and Related Subjects (DIS2026)\\
4 - 8 May 2026\\
Bologna, Italy\\}

\begin{document}
\maketitle

\section{Introduction}
Quarkonia, i.e.\ bound states of a heavy quark-antiquark pair ($Q\bar Q$, with $Q = c, b$) are key probes of QCD. The heavy-quark hierarchy $\MQ \gg \Lambda_{\rm QCD}$ provides a hard scale allowing for a perturbative treatment of the $Q\bar Q$ pair creation. On the one hand, their clean experimental signature makes them ideal tools to access the partonic structure of hadrons. On the other hand, their production mechanism is still an open issue. The three main models, 
(i) the Color Evaporation Model~\cite{Halzen:1977rs,Fritzsch:1977ay}, (ii) the Color-Singlet Model (CSM)~\cite{Baier:1983va} and (iii) the widely used effective-field-theory approach of nonrelativistic QCD (NRQCD)~\cite{Bodwin:1994jh}, all differ in describing the nonperturbative transition to the observable bound state. Even after decades of study, NRQCD is not able to consistently describe all cross-section and polarization measurements, and the long-distance matrix elements extracted at next-to-leading order remain affected by large uncertainties~\cite{Boer:2024ylx}.

Within the last decade, inclusive double-$\Jpsi$ production has been proposed as a way to probe the transverse momentum dependent (TMD) gluon distributions~\cite{Lansberg:2017dzg,Scarpa:2019fol}. For this process, the color-octet (CO) contributions in NRQCD are suppressed with respect to the color-singlet one by at least $\mathcal{O}(v^3)$~\cite{He:2015qya}, $v$ being the relative velocity of the $Q\bar Q$ pair. Therefore, the CSM is expected to be adequate and TMD factorization to be applicable. In fact, if the two $Q\bar Q$ pairs are produced in a colorless state, only initial-state interactions occur, in strong analogy with the Drell-Yan (DY) process. So far, all available analyses were performed for the gluon fusion channel $gg\to \Jpsi\,\Jpsi$. 

Motivated by the recent COMPASS measurement of di-$\Jpsi$ production in pion-nucleon scattering~\cite{COMPASS:2022djq} and by LHCb data in the collider mode~\cite{LHCb:2023ybt}, in Ref.~\cite{Flore:2025rmo} we studied for the first time the quark-antiquark annihilation channel, $\qqbar \to \mathcal{Q}\,\mathcal{Q}$, relevant at the lower center-of-mass energies of the present and future fixed-target facilities. In what follows we summarize our main results: the derivation of the angular structure of the cross section, and the phenomenological predictions for COMPASS/AMBER~\cite{Adams:2018pwt} and for the LHC fixed-target programs SMOG2~\cite{BoenteGarcia:2024kba} and LHCspin~\cite{LHCspin:2025lvj}.

\section{Formalism}
We consider the inclusive production of two quarkonium states (di-$\Jpsi$, di-$\psi(2S)$, di-$\Upsilon$) in inelastic collisions of two spin-$1/2$ hadrons
\begin{equation}
h_1(P_1,S_1) + h_2(P_2,S_2) \to \mathcal{Q}(K_1) + \mathcal{Q}(K_2) + X\,,
\end{equation}
with the two final-state quarkonia almost back-to-back in the transverse plane w.r.t.~the beams. Each quarkonium is considered in a $S$-wave, formed by a $Q\bar Q$ pair produced directly in the $^3S_1^{[1]}$ state. At leading order in $\alpha_s$ the underlying partonic channels are: (i) $\qqbar \to \mathcal{Q}\mathcal{Q}$ and (ii) $gg\to \mathcal{Q}\mathcal{Q}$. We focus on the former, whose scattering amplitude we computed in Ref.~\cite{Flore:2025rmo}. The corresponding unpolarized cross section, integrated over the azimuthal angles, agrees with the one obtained in Ref.~\cite{Kartvelishvili:1983lrw}.

To identify the TMD regime we introduce two combinations of transverse momenta, $\Kperp = (\bm{K}_{1\perp}-\bm{K}_{2\perp})/2$ and $\qT = \bm{K}_{1\perp}+\bm{K}_{2\perp}$, with $|\qT| \ll |\Kperp|$. Assuming TMD factorization, the cross section can be written as
\begin{multline}
\frac{{\rm d}\sigma}{{\rm d}y_1\,{\rm d}y_2\,{\rm d}^2\Kperp\,{\rm d}^2\qT} = \frac{1}{16\pi^2 s^2}\int {\rm d}^2\bm{k}_{1T}\,{\rm d}^2\bm{k}_{2T}\, \delta^2(\bm{k}_{1T}+\bm{k}_{2T}-\qT) \\ 
\times \sum_q \Big[ \Phi^q(x_1,\bm{k}_{1T}) \otimes \overline{\Phi}^q(x_2,\bm{k}_{2T}) \otimes |\mathcal{M}_{\qqbar\to\mathcal{Q}\mathcal{Q}}|^2 + \{\Phi^q \leftrightarrow \overline{\Phi}^q\}\Big]\,,
\label{eq:xsec}
\end{multline}
where $\Phi^q$ ($\overline\Phi^q$) is the quark (antiquark) TMD correlator, parametrized at leading twist in terms of the eight quark TMDs~\cite{Boer:1997nt}. In what follows, we express the results in terms of
\begin{equation}
z = \frac{K_1\cdot k_1}{k_1\cdot k_2} = \frac{1}{1+e^{y_1-y_2}}\,,\qquad M^2_{\mathcal{Q}\mathcal{Q}} = \frac{\Mperp^2}{z(1-z)}\,,\qquad Y_{\mathcal{Q}\mathcal{Q}} = \frac{y_1+y_2}{2}\,,
\end{equation}
with $y_1$ and $y_2$ being the rapidities of the two outgoing quarkonia in the center-of-mass frame of $h_1$ and $h_2$, whereas $\Mperp^2 = \MQ^2 + \Kperp^2$ is the quarkonium transverse mass.

In a frame where the longitudinal direction is set by the two incoming hadrons, and denoting by $\phi_T$, $\phi_\perp$, $\phi_{S_1}$, $\phi_{S_2}$ the azimuthal angles of $\qT$, $\Kperp$ and of the two initial-state hadron spins, the fully differential cross section can be expressed as
\begin{align}
{\rm d}\sigma ={}& \frac{131072}{243}\,\frac{\alpha_s^4}{\MQ^2 \Mperp^6 s}\,|R_{\mathcal Q}(0)|^4\,
z^4(1-z)^4 \bigg\{ F_{UU} + F_{UU}^{\cos 2(\phi_T-\phi_\perp)}\cos 2(\phi_T-\phi_\perp) \nonumber\\
&+ \Big[ S_{1L}\, F_{LU}^{\sin 2(\phi_T-\phi_\perp)} + S_{2L}\, F_{UL}^{\sin 2(\phi_T-\phi_\perp)} \Big]\sin 2(\phi_T-\phi_\perp)
+ |\bm{S}_{1T}|\Big[ F_{TU}^{\sin(\phi_T-\phi_{S_1})}\sin(\phi_T-\phi_{S_1}) \nonumber\\
&+ F_{TU}^{\sin(\phi_T+\phi_{S_1}-2\phi_\perp)}\sin(\phi_T+\phi_{S_1}-2\phi_\perp)
+ F_{TU}^{\sin(3\phi_T-\phi_{S_1}-2\phi_\perp)}\sin(3\phi_T-\phi_{S_1}-2\phi_\perp) \Big] \nonumber\\
&+ |\bm{S}_{2T}|\Big[ F_{UT}^{\sin(\phi_T-\phi_{S_2})}\sin(\phi_T-\phi_{S_2})
+ F_{UT}^{\sin(\phi_T+\phi_{S_2}-2\phi_\perp)}\sin(\phi_T+\phi_{S_2}-2\phi_\perp) \nonumber\\
&+ F_{UT}^{\sin(3\phi_T-\phi_{S_2}-2\phi_\perp)}\sin(3\phi_T-\phi_{S_2}-2\phi_\perp) \Big]
+ [\text{double polarized azimuthal modulations}] \bigg\}\,,
\label{eq:angular}
\end{align}
where $R_{\mathcal Q}(0)$ is the quarkonium radial wave function at the origin. The subscripts (superscripts) of the structure functions $F$ label the hadron polarizations (the azimuthal modulation). Each structure function factorizes into a hard part and a convolution of TMDs, 
\begin{align}
\Cc{wfg} \equiv
\int {\rm d}^2\bm{k}_{1T}\,{\rm d}^2\bm{k}_{2T}\, w(\bm{k}_{1T},\bm{k}_{2T})\,
f_{q/h_1}(x_1,\bm{k}_{1T})\, g_{\bar q/h_2}(x_2,\bm{k}_{2T})\,
\delta^2(\bm{k}_{1T}+\bm{k}_{2T}-\qT) \nonumber\\ \hfill + \{q\leftrightarrow\bar q\}\,,
\end{align}
where the function $w$ depends on $f_{q/h_1}$ and $g_{\bar q/h_2}$. It turns out that only two hard factors appear:
\begin{equation}
H \left(z,\tfrac{\MQ^2}{\Mperp^2}\right) = 5 - 12\,z(1-z) \left(1-\tfrac{\MQ^2}{\Mperp^2}\right) - \tfrac{\MQ^2}{\Mperp^2}\,,
\quad
\Delta H \left(z,\tfrac{\MQ^2}{\Mperp^2}\right) = - \left(1-\tfrac{\MQ^2}{\Mperp^2}\right) [1-12\,z(1-z)]\,,
\end{equation}
and that the azimuthal modulations and TMD convolutions are \emph{the same} as those obtained for the LO channel $\qqbar\to\gamma^*\to\ell^+\ell^-$ of the DY process~\cite{Arnold:2008kf}. 
This close correspondence, together with the DY-like color flow, makes the quark-induced channel an ideal complementary reaction to SIDIS and DY for the study of quark TMDs and their process dependence.

For brevity, below we focus on the unpolarized cross section and on the Sivers azimuthal asymmetries. We guide the reader to Ref.~\cite{Flore:2025rmo} for the azimuthal asymmetries involving the quark TMD transversity and Boer-Mulders functions. The Sivers asymmetry is given by~\cite{Flore:2025rmo}
\begin{equation}
\avg{\sin(\phi_T-\phi_S)} \equiv A_{UT}^{\sin(\phi_T-\phi_S)}
= \frac{F_{UT}^{\sin(\phi_T-\phi_S)}}{F_{UU}}
= \frac{\Cc{w^1_{UT}\,f_1^q\,f_{1T}^{\perp\,\bar q}}}{\Cc{f_1^q f_1^{\bar q}}}
\label{eq:sivers}
\end{equation}
and provides access to the quark Sivers function $\fperpT$ convoluted with the unpolarized TMD $f_1$. In what follows, as in the DY case, we assume the Sivers sign change w.r.t.~SIDIS.

\section{Results}
For our predictions we adopt recent TMD parametrizations from the MAP Collaboration through {\tt TMDlib}~\cite{Hautmann:2014kza,Abdulov:2021ivr}: the MAPTMDPion22 set for the pion~\cite{Rossi:private}, the PV17~\cite{Bacchetta:2017gcc} and MAP22 sets for the proton~\cite{Cerutti:2022lmb}, and the PV20 quark Sivers set~\cite{Bacchetta:2020gko}, all at next-to-leading-logarithmic (NLL) accuracy. The choice $\lvert R_{\cal Q}(0)\rvert^2 = 1.0$ GeV$^3$ comes from a power-law potential. 
The hard scale is taken to be equal to the invariant mass of the pair, $Q = M_{\mathcal{Q}\mathcal{Q}}$, and single-parton scattering is assumed to dominate. By using \texttt{HELAC-Onia}~\cite{Shao:2015vga}, we explicitly checked that the CS-CO background is at most $\sim 10\%$ of the CS-CS channel, confirming the expected NRQCD suppression.

\paragraph{Di-$\Jpsi$ at COMPASS}

\begin{figure}[t]
\centering
\includegraphics[height=5.15cm, keepaspectratio]{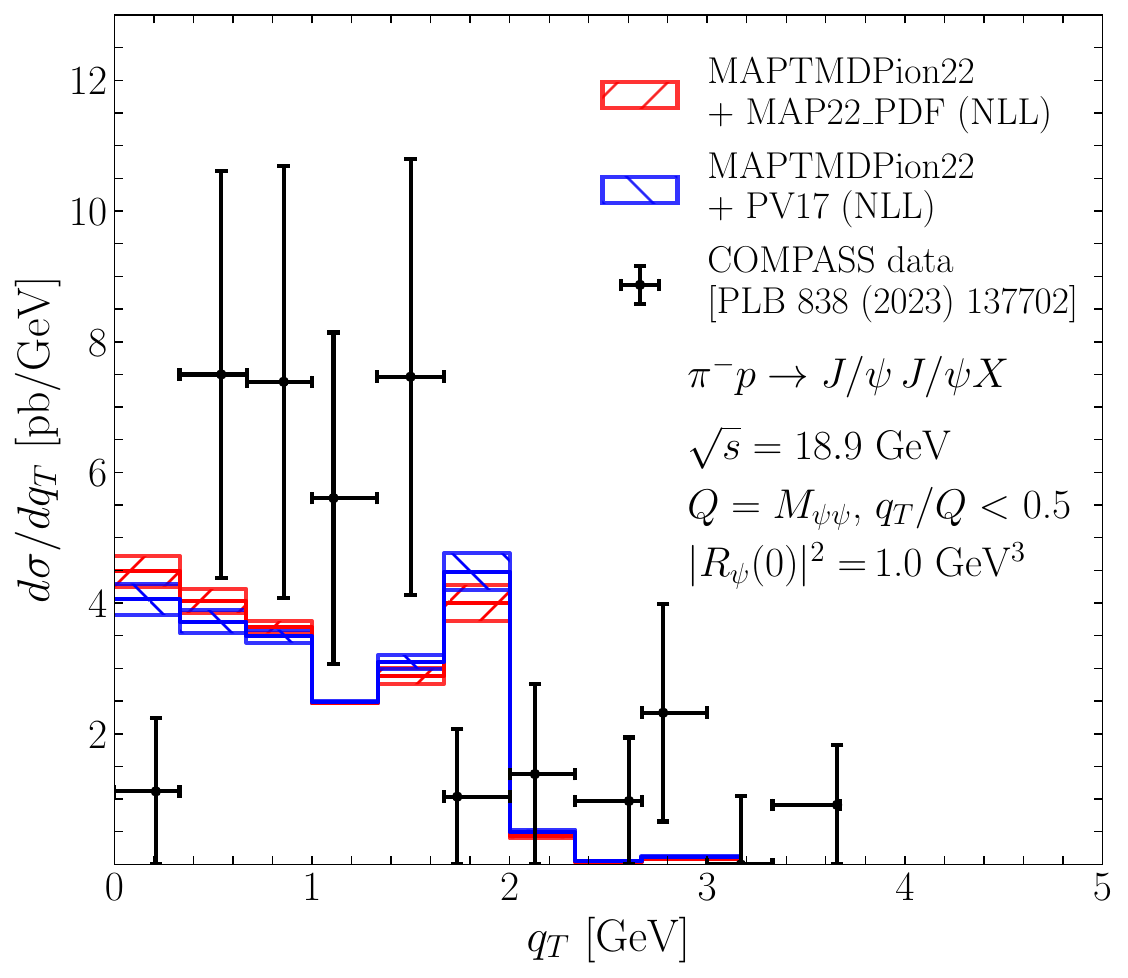}
\includegraphics[height=5.15cm, keepaspectratio]{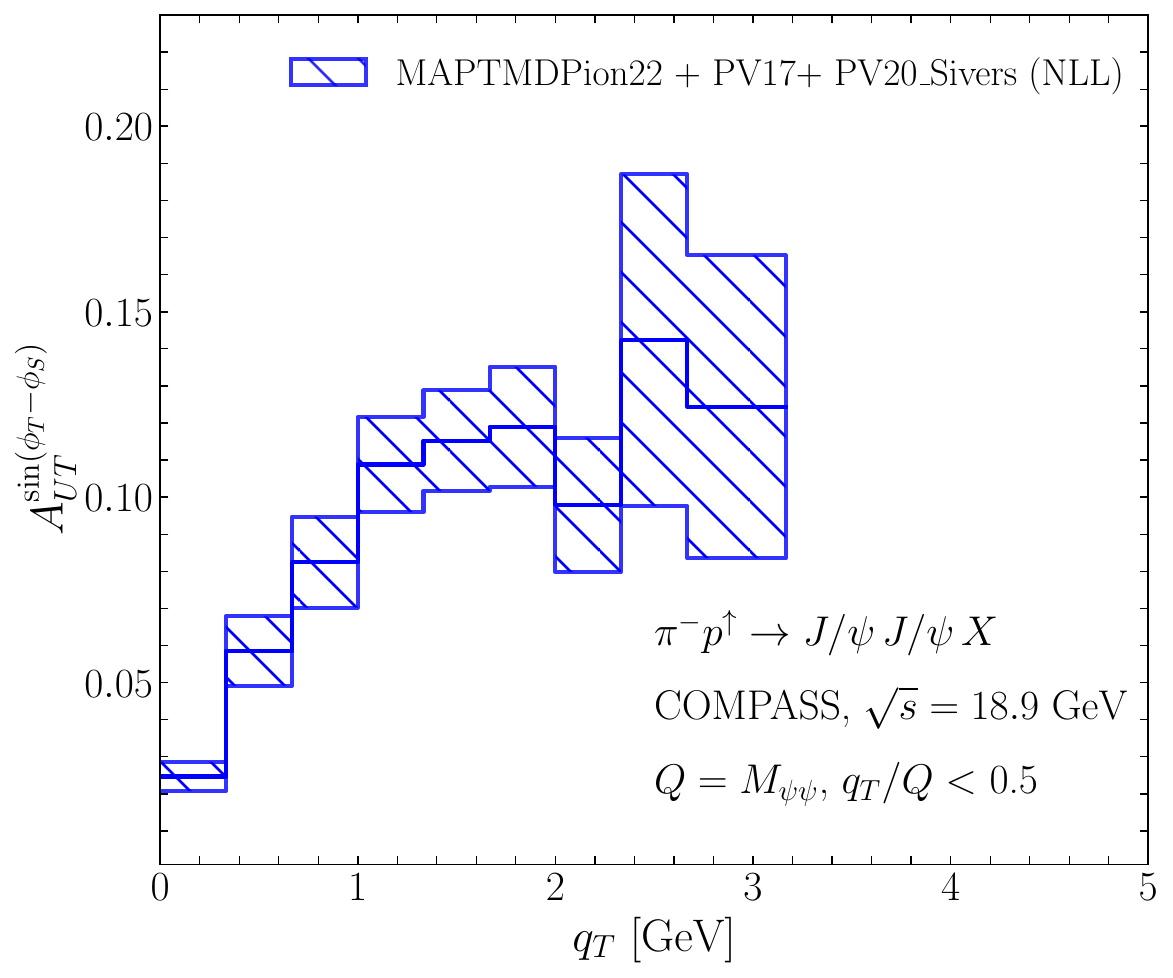}
\caption{Left: comparison with COMPASS data~\cite{COMPASS:2022djq} of the unpolarized cross section for di-$\Jpsi$ production in $\pi^-p$ scattering, as a function of the transverse momentum of the $\Jpsi$-pair, $q_T$, at $\sqrts = 18.9$~GeV. Right: prediction for the Sivers asymmetry $A_{UT}^{\sin(\phi_T-\phi_S)}$ in $\pi^-p^\uparrow$ collisions at the same energy, assuming the Sivers sign change and adopting the same $q_T$-binning as for the unpolarized cross section.}
\label{fig:compass}
\end{figure}

COMPASS measured the unpolarized cross section for $\pi^- p \to \Jpsi\,\Jpsi\,X$ using a $190$~GeV pion beam, at $\sqrts \simeq 18.9$~GeV. Here, double-parton scattering is negligible, being at most at the level of 8\%~\cite{COMPASS:2022djq}. In Fig.~\ref{fig:compass} (left), we present the comparison of our predictions based on the MAP22 and PV17 sets with the COMPASS data. The gluon-induced channel, computed following Ref.~\cite{Bor:2025ztq}, turns out to be negligible ($\mathcal{O}(10^{-3})$~pb). The agreement with data is fairly good given the limited statistics ($\sim 25$ events). Although next-to-leading-order QCD corrections and feed-down contributions could refine and improve the description, the present accuracy is adequate as a baseline for the azimuthal asymmetries, in which the TMD uncertainties partially cancel in the ratios.

In Fig.~\ref{fig:compass} (right)  we present predictions for the Sivers asymmetry in Eq.~\eqref{eq:sivers}. A sizable Sivers asymmetry, up to $10$--$15\%$, is obtained. The sign of the asymmetry is driven by the dominant $\bar u_\pi u_p$ channel, for which the valence region is probed in both the pion and the transversely polarized proton. Repeating such a measurement at AMBER would significantly improve the current statistical precision.
\vspace*{-1mm}
\paragraph{Di-quarkonium production at SMOG2 and LHCspin}

\begin{figure}[htbp]
\centering
\includegraphics[width=0.32\textwidth,keepaspectratio]{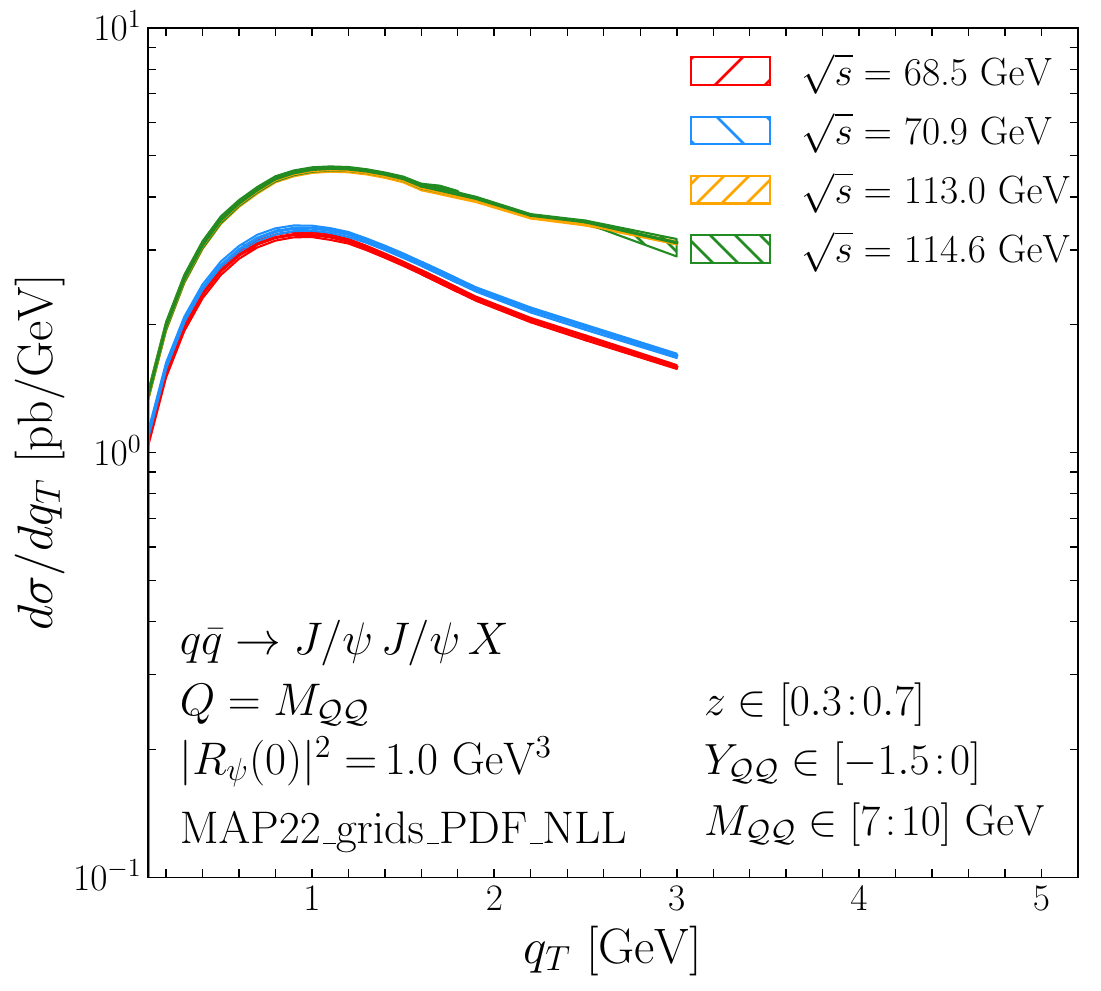}
\includegraphics[width=0.32\textwidth,keepaspectratio]{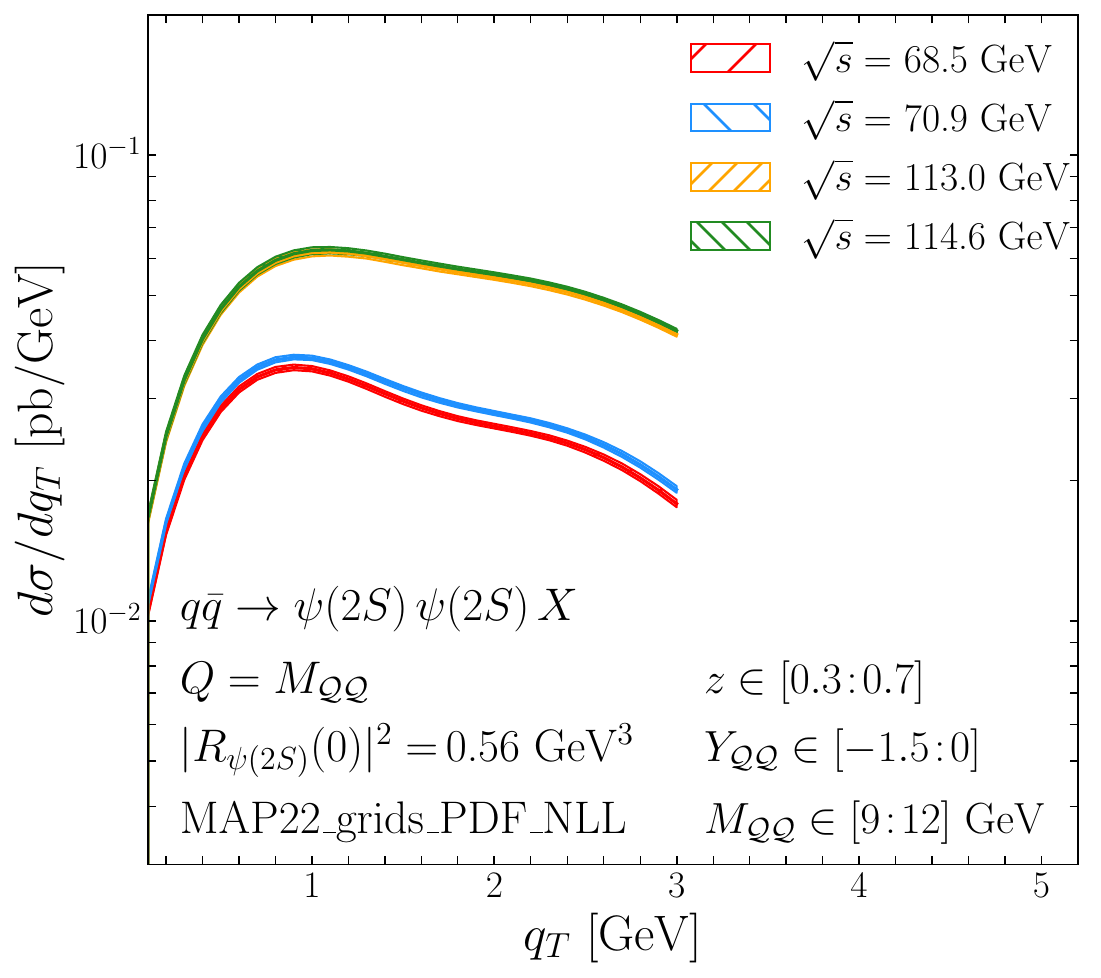}
\includegraphics[width=0.32\textwidth,keepaspectratio]{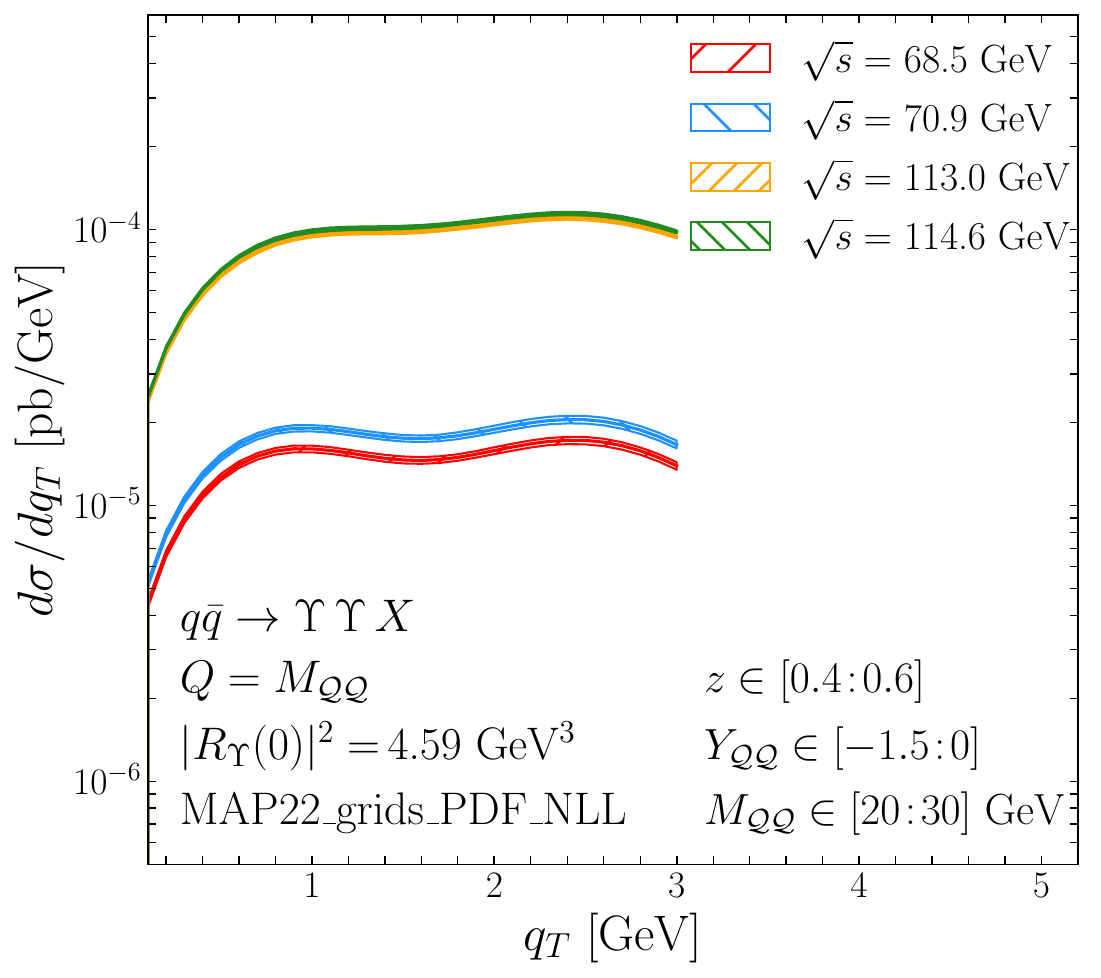}\\[4pt]
\includegraphics[width=0.32\textwidth,keepaspectratio]{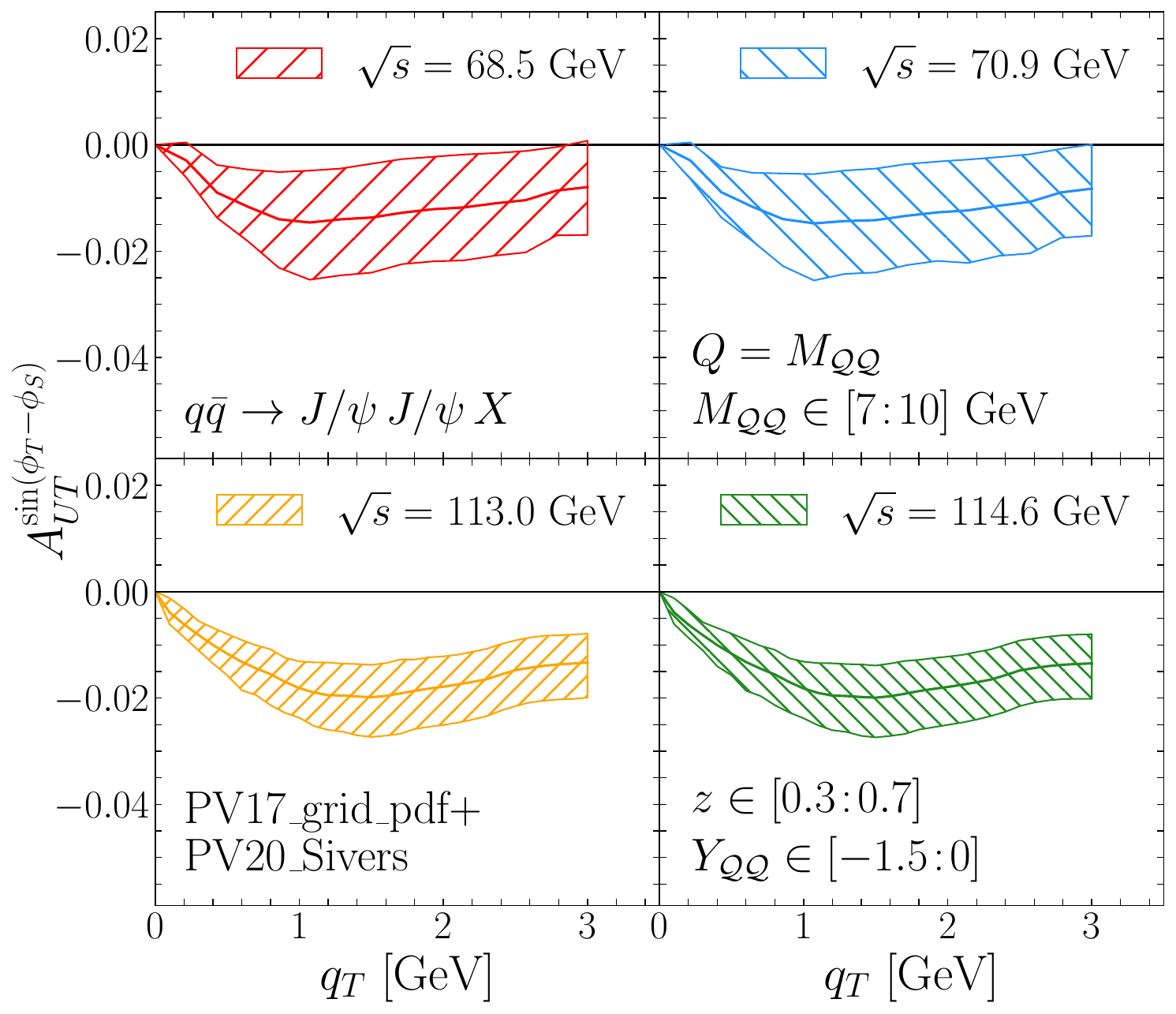}
\includegraphics[width=0.32\textwidth,keepaspectratio]{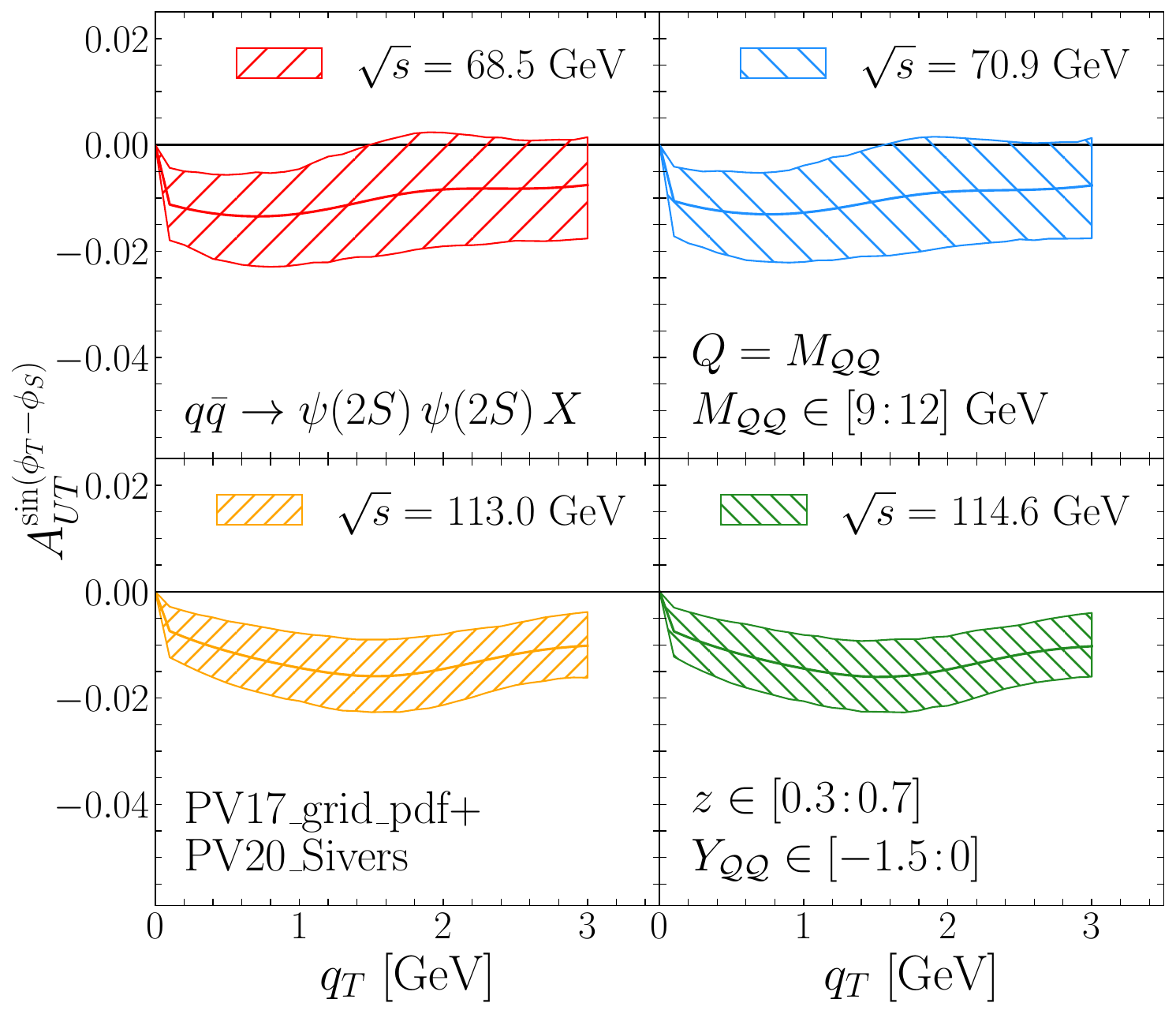}
\includegraphics[width=0.32\textwidth,keepaspectratio]{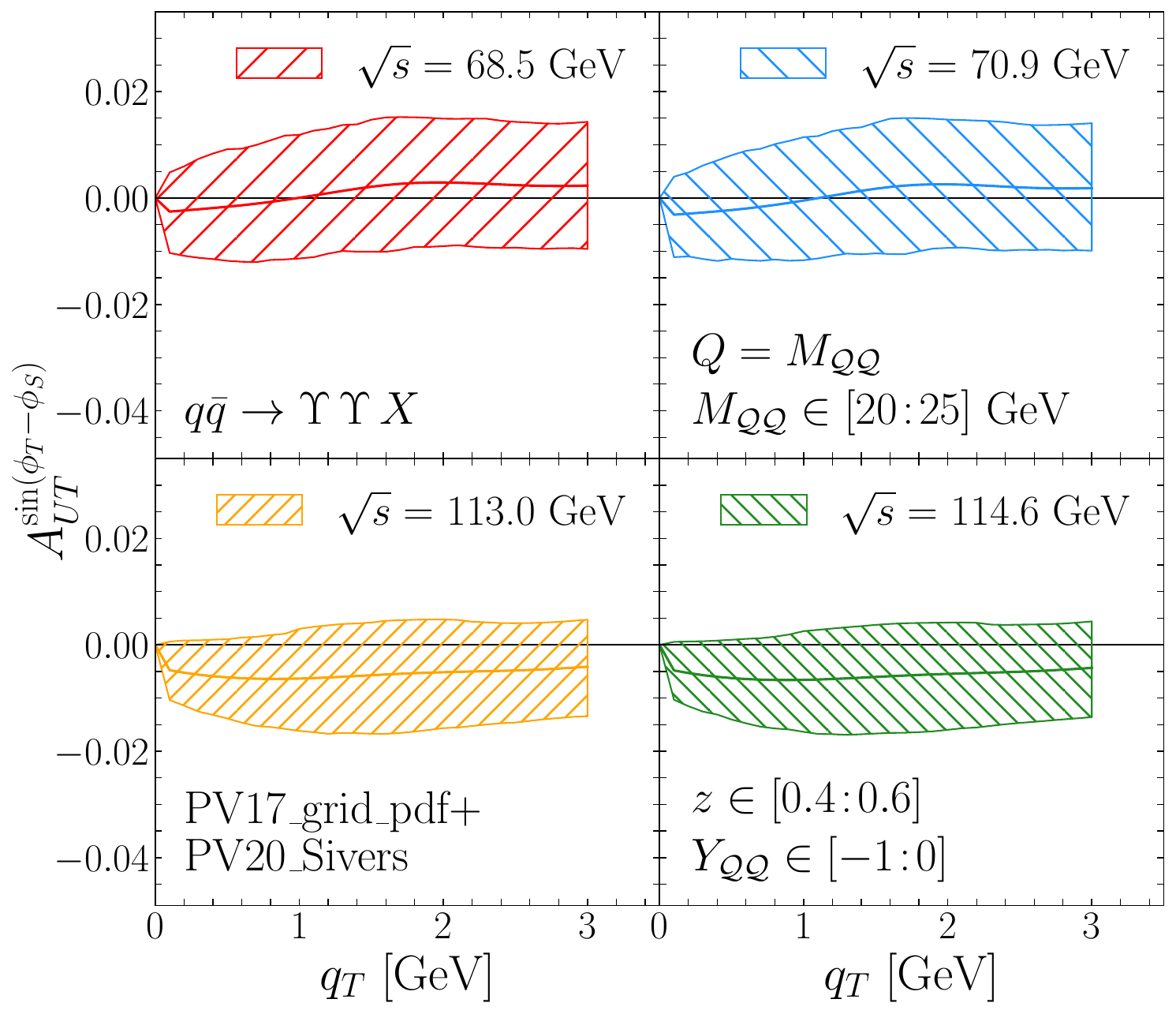}
\caption{Predictions for double quarkonium production (di-$\Jpsi$, di-$\psi(2S)$ and di-$\Upsilon$) at the LHC fixed-target experiments, for different values of $\sqrts$ and in different $Y_{\mathcal{Q}\mathcal{Q}}$, $z$ and $M_{\mathcal{Q}\mathcal{Q}}$ ranges. Top: quark-induced unpolarized cross section based on the MAP22 TMDs. Bottom: quark-induced Sivers asymmetry, assuming the Sivers sign change w.r.t.~SIDIS.}
\label{fig:smog}
\end{figure}

In Fig.~\ref{fig:smog} (top) we present predictions for the LHC fixed-target programs for di-$\Jpsi$, di-$\psi(2S)$ and di-$\Upsilon$ production as a function of $q_T$ in selected ranges of $M_{\mathcal{Q}\mathcal{Q}}$, $Y_{\mathcal{Q}\mathcal{Q}}$ and $z$. At variance with COMPASS, the gluon fusion channel is no longer negligible, being $30$--$40\%$ of the $\qqbar$ one at $\sqrts \approx 70$~GeV and about a factor of two larger at $\sqrts \approx 115$~GeV. 
With the projected Run~3 luminosity the di-$\Jpsi$ yield is expected to be measurable at LHCb in the fixed-target mode, with similar expectations for di-$\psi(2S)$ and di-$\Upsilon$. The latter would be a valuable tool to constrain the poorly known unpolarized distributions in the large-$x$ region.

The corresponding Sivers asymmetries are presented in Fig.~\ref{fig:smog} (bottom).  At variance with COMPASS, the asymmetry is negative. This follows from the different flavor composition of $\pi^-$ and $p$: in $pp$ collisions the $d$-quark and sea-quark Sivers functions, coupled to valence quarks of the unpolarized proton, contribute with opposite sign w.r.t.~the $u$-quark one. A small, negative asymmetry of $1-2\%$ is predicted for di-$\Jpsi$ and di-$\psi(2S)$, while for di-$\Upsilon$ it is compatible with zero because of the different $x_2$ range probed. A larger measured asymmetry would therefore signal a nonzero gluon Sivers function. Dedicated studies at LHCspin can thus constrain the quark Sivers function in a kinematic region complementary to SIDIS.

\section{Conclusions}
We have studied double quarkonium production in unpolarized and polarized hadron-hadron collisions at fixed-target experiments within TMD factorization and the CSM. The quark-antiquark annihilation channel exhibits the same azimuthal modulations and color-flow structure as the DY process. At COMPASS energies the gluon contribution is strongly suppressed, making these measurements ideal to probe quark TMDs. Based on recent TMDs extractions, we predicted a sizable Sivers asymmetry of $10$--$15\%$, dominated by the $\bar u_\pi u_p$ channel. At the LHC fixed-target experiments the $\qqbar$ channel turns out to be dominant only up to $\sqrts \sim 70$~GeV. There the Sivers asymmetries are small ($1$--$2\%$ for di-$\Jpsi$ and di-$\psi(2S)$, even smaller for di-$\Upsilon$), and the process would therefore be an important probe of gluon TMDs. 

\acknowledgments
C.F.\ is supported by the European Union's Horizon Europe research and innovation programme under the Marie Sk\l odowska-Curie grant agreement n.~101150792 (STAT-TMDs). The work of C.P. is supported by Fondazione di Sardegna through the project "Studies of charmonia and bottomonia to understand quarkonium formation and transverse momentum dependent distributions in both collider and fixed-target modes at the Large Hadron Collider at CERN)", No. F83C26000350007.

\bibliographystyle{JHEP}
\setlength{\bibsep}{\itemsep}
\bibliography{references}

@article{Bodwin:1994jh,
    author = "Bodwin, Geoffrey T. and Braaten, Eric and Lepage, G. Peter",
    title = "{Rigorous QCD analysis of inclusive annihilation and production of heavy quarkonium}",
    eprint = "hep-ph/9407339",
    archivePrefix = "arXiv",
    reportNumber = "ANL-HEP-PR-94-24, FERMILAB-PUB-94-073-T, NUHEP-TH-94-5",
    doi = "10.1103/PhysRevD.55.5853",
    journal = "Phys. Rev. D",
    volume = "51",
    pages = "1125--1171",
    year = "1995",
    note = "[Erratum: Phys.Rev.D 55, 5853 (1997)]"
}

@article{Baier:1983va,
    author = "Baier, R. and Ruckl, R.",
    title = "{Hadronic Collisions: A Quarkonium Factory}",
    reportNumber = "BI-TP 83/02",
    doi = "10.1007/BF01572254",
    journal = "Z. Phys. C",
    volume = "19",
    pages = "251",
    year = "1983"
}

@article{Boer:2024ylx,
    author = {Boer, Dani{\"e}l and others},
    title = "{Physics case for quarkonium studies at the Electron Ion Collider}",
    eprint = "2409.03691",
    archivePrefix = "arXiv",
    primaryClass = "hep-ph",
    doi = "10.1016/j.ppnp.2025.104162",
    journal = "Prog. Part. Nucl. Phys.",
    volume = "142",
    pages = "104162",
    year = "2025"
}

@phdthesis{Bor:2025ztq,
    author = "Bor, Jelle",
    title = "{Gluon-induced quarkonium production in transverse-momentum-dependent factorisation: applications to the LHC and EIC}",
    reportNumber = "tel-04968266, 2025UPASP010",
    doi = "10.33612/diss.1191891247",
    school = "U. Groningen, VSI, U. Paris-Saclay",
    year = "2025"
}

@article{LHCb:2023ybt,
    author = "Aaij, Roel and others",
    collaboration = "LHCb",
    title = "{Measurement of J/{\ensuremath{\psi}}-pair production in pp collisions at $ \sqrt{s} $ = 13 TeV and study of gluon transverse-momentum dependent PDFs}",
    eprint = "2311.14085",
    archivePrefix = "arXiv",
    primaryClass = "hep-ex",
    reportNumber = "LHCb-PAPER-2023-022, CERN-EP-2023-242",
    doi = "10.1007/JHEP03(2024)088",
    journal = "JHEP",
    volume = "03",
    pages = "088",
    year = "2024"
}

@article{COMPASS:2022djq,
    author = "Alexeev, G. D. and others",
    collaboration = "COMPASS",
    title = "{Double J/\ensuremath{\psi} production in pion-nucleon scattering at COMPASS}",
    eprint = "2204.01817",
    archivePrefix = "arXiv",
    primaryClass = "hep-ex",
    reportNumber = "CERN-EP-2022-073",
    doi = "10.1016/j.physletb.2023.137702",
    journal = "Phys. Lett. B",
    volume = "838",
    pages = "137702",
    year = "2023"
}

@article{Arnold:2008kf,
    author = "Arnold, S. and Metz, A. and Schlegel, M.",
    title = "{Dilepton production from polarized hadron hadron collisions}",
    eprint = "0809.2262",
    archivePrefix = "arXiv",
    primaryClass = "hep-ph",
    reportNumber = "JLAB-THY-08-877",
    doi = "10.1103/PhysRevD.79.034005",
    journal = "Phys. Rev. D",
    volume = "79",
    pages = "034005",
    year = "2009"
}

@article{Hautmann:2014kza,
    author = {Hautmann, F. and Jung, H. and Kr\"amer, M. and Mulders, P. J. and Nocera, E. R. and Rogers, T. C. and Signori, A.},
    title = "{TMDlib and TMDplotter: library and plotting tools for transverse-momentum-dependent parton distributions}",
    eprint = "1408.3015",
    archivePrefix = "arXiv",
    primaryClass = "hep-ph",
    reportNumber = "DESY-14-059, NIKHEF-2014-024, YITP-SB-14-24",
    doi = "10.1140/epjc/s10052-014-3220-9",
    journal = "Eur. Phys. J. C",
    volume = "74",
    pages = "3220",
    year = "2014"
}

@article{Abdulov:2021ivr,
    author = "Abdulov, N. A. and others",
    title = "{TMDlib2 and TMDplotter: a platform for 3D hadron structure studies}",
    eprint = "2103.09741",
    archivePrefix = "arXiv",
    primaryClass = "hep-ph",
    reportNumber = "DESY 21-026, DESY-21-026, IFJPAN-IV-2021-4, JLAB-THY-21-3337",
    doi = "10.1140/epjc/s10052-021-09508-8",
    journal = "Eur. Phys. J. C",
    volume = "81",
    number = "8",
    pages = "752",
    year = "2021"
}

@misc{Rossi:private,
    author = "{L. Rossi}",
    howpublished = "private communication",
    year = "2025"
}

@article{Cerutti:2022lmb,
    author = "Cerutti, Matteo and Rossi, Lorenzo and Venturini, Simone and Bacchetta, Alessandro and Bertone, Valerio and Bissolotti, Chiara and Radici, Marco",
    collaboration = "MAP (Multi-dimensional Analyses of Partonic distributions)",
    title = "{Extraction of pion transverse momentum distributions from Drell-Yan data}",
    eprint = "2210.01733",
    archivePrefix = "arXiv",
    primaryClass = "hep-ph",
    doi = "10.1103/PhysRevD.107.014014",
    journal = "Phys. Rev. D",
    volume = "107",
    number = "1",
    pages = "014014",
    year = "2023"
}

@article{Bacchetta:2017gcc,
    author = "Bacchetta, Alessandro and Delcarro, Filippo and Pisano, Cristian and Radici, Marco and Signori, Andrea",
    title = "{Extraction of partonic transverse momentum distributions from semi-inclusive deep-inelastic scattering, Drell-Yan and Z-boson production}",
    eprint = "1703.10157",
    archivePrefix = "arXiv",
    primaryClass = "hep-ph",
    reportNumber = "JLAB-THY-17-2437",
    doi = "10.1007/JHEP06(2017)081",
    journal = "JHEP",
    volume = "06",
    pages = "081",
    year = "2017",
    note = "[Erratum: JHEP 06, 051 (2019)]"
}

@article{Bacchetta:2020gko,
    author = "Bacchetta, Alessandro and Delcarro, Filippo and Pisano, Cristian and Radici, Marco",
    title = "{The 3-dimensional distribution of quarks in momentum space}",
    eprint = "2004.14278",
    archivePrefix = "arXiv",
    primaryClass = "hep-ph",
    reportNumber = "JLAB-THY-20-3186",
    doi = "10.1016/j.physletb.2022.136961",
    journal = "Phys. Lett. B",
    volume = "827",
    pages = "136961",
    year = "2022"
}

@article{LHCspin:2025lvj,
    author = "Accardi, A. and others",
    collaboration = "LHCspin",
    title = "{LHCspin: a Polarized Gas Target for LHC}",
    eprint = "2504.16034",
    archivePrefix = "arXiv",
    primaryClass = "hep-ex",
    reportNumber = "FERMILAB-PUB-25-0295-PPD",
    month = "4",
    year = "2025"
}

@article{BoenteGarcia:2024kba,
    author = "Boente Garcia, O. and others",
    title = "{High-density gas target at the LHCb experiment}",
    eprint = "2407.14200",
    archivePrefix = "arXiv",
    primaryClass = "physics.ins-det",
    reportNumber = "LHCb-DP-2024-002",
    doi = "10.1103/PhysRevAccelBeams.27.111001",
    journal = "Phys. Rev. Accel. Beams",
    volume = "27",
    number = "11",
    pages = "111001",
    year = "2024"
}

@article{Boer:1997nt,
    author = "Boer, Daniel and Mulders, P. J.",
    title = "{Time reversal odd distribution functions in leptoproduction}",
    eprint = "hep-ph/9711485",
    archivePrefix = "arXiv",
    reportNumber = "NIKHEF-97-049, VUTH-97-20",
    doi = "10.1103/PhysRevD.57.5780",
    journal = "Phys. Rev. D",
    volume = "57",
    pages = "5780--5786",
    year = "1998"
}

@article{Lansberg:2017dzg,
    author = "Lansberg, Jean-Philippe and Pisano, Cristian and Scarpa, Florent and Schlegel, Marc",
    title = "{Pinning down the linearly-polarised gluons inside unpolarised protons using quarkonium-pair production at the LHC}",
    eprint = "1710.01684",
    archivePrefix = "arXiv",
    primaryClass = "hep-ph",
    doi = "10.1016/j.physletb.2018.08.004",
    journal = "Phys. Lett. B",
    volume = "784",
    pages = "217--222",
    year = "2018",
    note = "[Erratum: Phys.Lett.B 791, 420--421 (2019)]"
}

@article{Scarpa:2019fol,
    author = {Scarpa, Florent and Boer, Dani\"el and Echevarria, Miguel G. and Lansberg, Jean-Philippe and Pisano, Cristian and Schlegel, Marc},
    title = "{Studies of gluon TMDs and their evolution using quarkonium-pair production at the LHC}",
    eprint = "1909.05769",
    archivePrefix = "arXiv",
    primaryClass = "hep-ph",
    doi = "10.1140/epjc/s10052-020-7619-1",
    journal = "Eur. Phys. J. C",
    volume = "80",
    number = "2",
    pages = "87",
    year = "2020"
}

@article{Adams:2018pwt,
    author = "Adams, B. and others",
    title = "{Letter of Intent: A New QCD facility at the M2 beam line of the CERN SPS (COMPASS++/AMBER)}",
    eprint = "1808.00848",
    archivePrefix = "arXiv",
    primaryClass = "hep-ex",
    reportNumber = "CERN-SPSC-2019-003, SPSC-I-250",
    month = "8",
    year = "2018"
}

@article{Kartvelishvili:1983lrw,
    author = "Kartvelishvili, V. G. and Esakiya, Sh. M.",
    title = "{ON HADRON INDUCED PRODUCTION OF J / PSI MESON PAIRS. (IN RUSSIAN)}",
    journal = "Yad. Fiz.",
    volume = "38",
    pages = "722--726",
    year = "1983"
}

@article{He:2015qya,
    author = "He, Zhi-Guo and Kniehl, Bernd A.",
    title = "{Complete Nonrelativistic-QCD Prediction for Prompt Double J/{\ensuremath{\psi}} Hadroproduction}",
    eprint = "1609.02786",
    archivePrefix = "arXiv",
    primaryClass = "hep-ph",
    reportNumber = "DESY-15-104",
    doi = "10.1103/PhysRevLett.115.022002",
    journal = "Phys. Rev. Lett.",
    volume = "115",
    number = "2",
    pages = "022002",
    year = "2015"
}

@article{Shao:2015vga,
    author = "Shao, Hua-Sheng",
    title = "{HELAC-Onia 2.0: an upgraded matrix-element and event generator for heavy quarkonium physics}",
    eprint = "1507.03435",
    archivePrefix = "arXiv",
    primaryClass = "hep-ph",
    reportNumber = "CERN-PH-TH-2015-155",
    doi = "10.1016/j.cpc.2015.09.011",
    journal = "Comput. Phys. Commun.",
    volume = "198",
    pages = "238--259",
    year = "2016"
}

@article{Flore:2025rmo,
    author = "Flore, Carlo and Pisano, Cristian",
    title = "{Quark polarization and transverse momentum effects on double quarkonium production in hadronic collisions}",
    eprint = "2508.15482",
    archivePrefix = "arXiv",
    primaryClass = "hep-ph",
    doi = "10.1007/JHEP02(2026)085",
    journal = "JHEP",
    volume = "02",
    pages = "085",
    year = "2026"
}

@article{Halzen:1977rs,
    author = "Halzen, F.",
    title = "{Cvc for Gluons and Hadroproduction of Quark Flavors}",
    reportNumber = "Print-77-0411 (RUTHERFORD), RL-77-050/A",
    doi = "10.1016/0370-2693(77)90144-7",
    journal = "Phys. Lett. B",
    volume = "69",
    pages = "105--108",
    year = "1977"
}

@article{Fritzsch:1977ay,
    author = "Fritzsch, Harald",
    title = "{Producing Heavy Quark Flavors in Hadronic Collisions: A Test of Quantum Chromodynamics}",
    reportNumber = "CALT-68-582",
    doi = "10.1016/0370-2693(77)90108-3",
    journal = "Phys. Lett. B",
    volume = "67",
    pages = "217--221",
    year = "1977"
}

\end{document}